\documentclass[preprint]{aastex6}

\usepackage{graphicx}
\usepackage{natbib}
\usepackage[english]{babel}  
\usepackage{subfigure}
\usepackage{url}
\newcommand{\new}{ }
\newcommand{\newb}{ }
\newcommand{\neww}{ }
\newcommand{\newc}{  }
\newcommand{\newd}{ }
\newcommand{\newe}{ }

\begin{document}

\title{Large-amplitude quasi-periodic pulsations as evidence of impulsive heating  in hot transient loop systems detected in the EUV with SDO/AIA}



\author{Fabio Reale\altaffilmark{1}}
\affiliation{Dipartimento di Fisica \& Chimica, Universit\`a di Palermo,
              Piazza del Parlamento 1, 90134 Palermo, Italy;
              fabio.reale@unipa.it}
\author{Paola Testa}
\affiliation{Harvard -- Smithsonian Center for Astrophysics, 60 Garden St., Cambridge, MA 02138, USA}
\author{Antonino Petralia}
\affiliation{INAF-Osservatorio Astronomico di Palermo, Piazza del Parlamento 1, 90134 Palermo, Italy}
\author{Dmitrii Y. Kolotkov}
\affiliation{Centre for Fusion, Space and Astrophysics, Department of Physics, University of Warwick, Coventry CV4 7AL, UK}

\altaffiltext{1}{and INAF-Osservatorio Astronomico di Palermo, Piazza del Parlamento 1, 90134 Palermo, Italy}

\begin{abstract}
Short heat pulses can trigger plasma \neww{pressure fronts} inside closed magnetic tubes in the corona. The alternation of condensations and rarefactions \neww{from the pressure modes} drive large-amplitude pulsations in the plasma emission. Here we show the detection of such pulsations along magnetic tubes that brighten transiently in the hot 94~\AA\ EUV channel of SDO/AIA. The pulsations are consistent with those predicted by hydrodynamic loop modeling, and confirm pulsed heating in the loop system. The comparison of observations and model provides constraints on the heat deposition: a good agreement requires \new{loop twisting} and pulses deposited close to the footpoints with a duration of 0.5~min in one loop, and deposited in the corona with a duration of 2.5~min in another loop of the same loop system.
\end{abstract}


\section{Introduction} 
\label{sec:intro}

The diagnostic of coronal heating \new{both in quiescent and flaring conditions} is an important issue \citep[e.g.,][]{Klimchuk2006a,Reale2014a}. Several mechanisms have been invoked, from waves propagating upwards from the chromosphere \citep[e.g.,][]{Ionson1978a,Ofman1998a,van-Ballegooijen2011a,McIntosh2011a,Morton2019a} to small-scale reconnection in the corona \citep[e.g.,][]{Parker1988a,Kopp1993a,Vekstein2000a,Hood2009b,Priest2011a,Cargill2015a}, but unique signatures are difficult to find. Indirect evidence can therefore be very important. 

\neww{Flare-like rapid enhancements of the coronal emission are often observed to be accompanied by quasi-periodic pulsations (QPP), manifested as relatively short-lived modulations of the light curves by wave and oscillatory processes in active regions \citep[see reviews by][]{Nakariakov2009a,Van-Doorsselaere2016a}. Such pulsations are seen to be omnipresent in both solar \citep[e.g.,][]{Kupriyanova2010a,Simoes2015a,Inglis2016a} and stellar \citep[e.g.,][]{Cho2016a,Pugh2016a,Doyle2018a} emission throughout the whole electromagnetic spectrum and in both thermal and non-thermal emissions. One can distinguish at least three groups of potential physical mechanisms which could be responsible for the observed QPP \citep[see][for the most recent detailed review of the QPP mechanisms]{McLaughlin2018a}: direct modulation of the emitting plasma parameters by eigen magnetohydrodynamic (MHD) modes of an oscillating loop \citep[e.g.,][]{Nakariakov2005a,De-Moortel2012a}, periodic triggering of the energy release processes by external MHD oscillations \citep[e.g.,][]{Nakariakov2006a,McLaughlin2011b}, and regimes of repetitive energy releases ongoing in a self-oscillatory manner \citep[e.g.,][]{Kliem2000a,Murray2009a,Thurgood2019a}.
Rapidly decaying pulsations detected in thermal emission, with periods of several minutes  and with relatively small amplitudes ($\leq 10$\%) were generally interpreted in terms of compressive slow magnetoacoustic waves \citep[e.g.,][]{Kim2012a,Kolotkov2018a}, with exceptions \citep[e.g.,][]{Svestka1994a}. Those waves have been  detected in hotter UV and EUV channels, e.g., in flare-like events in Fe\,{\sc xix} 1118~\AA\ and Fe\,{\sc xxi} 1354~\AA\ spectral line profiles with the SUMER spectrograph on SOHO \citep{Wang2011a}, and more recently, in the SDO/AIA 131~\AA\ and 94~\AA\ channels \citep{Kumar2013a,Kumar2015a,Wang2015a}, and they are typically interpreted as magnetoacoustic wavefronts that evolve into slow standing waves. Slow waves, modulating the observed thermal emission intensity via density and temperature perturbations,  
have been studied extensively through hydrodynamic and MHD modeling of plasma inside closed magnetic tubes. Much attention has been devoted to wave generation by upflows \citep{Selwa2005a,Fang2015a} or velocity drivers \citep{Wang2013a} and damping and dissipation of harmonics through thermal conduction and viscosity \citep{Ofman2002a,De-Moortel2003a,Wang2018a}, gravitational stratification and magnetic field line divergence \citep{De-Moortel2004a}, nonlinear damping \citep{Verwichte2008a,Ruderman2013a}, mode coupling \citep{De-Moortel2004b}, and their relative efficiency in specific physical conditions.
More recent works by \cite{Wang2015a} and \cite{Wang2018a} showed the need for the modification of the thermal conduction and viscosity coefficients from their standard estimates (suppressed or enhanced by up to an order of magnitude) in order to match the observational properties of slow waves. This suggests that neither conduction nor viscosity is always a sufficient mechanism for the interpretation of the observed behaviour of slow waves, and additional physical effects should be taken into account. For example, a wave-induced perturbation of the thermodynamical equilibrium between some unspecified plasma heating and radiative cooling was shown to strongly affect the dynamics of slow waves, leading to their enhanced damping or amplification \citep{Kumar2016a,Nakariakov2017a}. Moreover, recent statistical study of slow wave properties \citep{Nakariakov2019a} revealed scaling laws between the oscillation periods, damping times, amplitudes, and the loop temperature, that require further theoretical interpretation. In particular, the damping of spatial harmonics of the sloshing (i.e., propagating and reflecting slow oscillations) was observed to be different to that in the standing slow waves. As such, slow coronal magnetoacoustic waves and a related modulation of the coronal emission in the form of QPP remain an actively developing and interesting research avenue, providing important insights into the field of MHD coronal seismology, contributing, in particular, to probing coronal plasma transport coefficients and constraining properties of the coronal heating function.}

\neww{For example, an advanced recent, time-dependent hydrodynamic loop modeling that includes all relevant physical effects (gravity, thermal conduction, radiative losses, plasma heating) and a highly stratified atmosphere from the chromosphere to the corona shows that short heat pulses can easily trigger significant pressure fronts that move back and forth along a loop and settle down to low-order standing modes with long wavelength \citep{Reale2016a}. } The heat pulses must be so short-lasting that the pressure does not have enough time to reach an equilibrium inside the loop during the heating phase and the sudden switch off creates a localized pressure drop which drives the \neww{fronts}. \neww{The efficient thermal conduction smooths out temperature perturbations in the hot corona, but the pressure fronts drive strong density fronts} that periodically accumulate when they are reflected and in turn determine strong alternating excesses and reductions of emission measure which modulate the light curves with long-period and large-amplitude pulsations. Hydrodynamic loop modeling has been able to 
reproduce long flares showing pulsations with period of hours and amplitude of $\sim 20$\%  observed in the X-rays from protostars in the Orion region \citep{Reale2018a}. 

\new{On the Sun, bright UV spots} detected in active regions with the Interface Region Imaging Spectrograph (IRIS) at the footpoints of transient hot loops could be explained with the presence of pulsed energy release in the corona \citep[][Testa et al. 2019, in preparation]{Testa2014a,Polito2018a}. Detected Doppler shifts were modelled with the release of non-thermal electron beams. It has been shown that these hot spots are at the footpoints of coronal loops that brighten for few minutes in the hottest EUV SDO/AIA channels, thus involving plasma heated to more than 10~MK \citep{Reale2019a}.

In these \neww{coronal loops} we may detect pulsations. Full-disk monitoring with the Atmospheric Imaging Assembly (AIA) on-board the Solar Dynamics Observatory allows for a full temporal and spatial coverage. The image cadence ($\sim 10$~s) is sufficiently high to resolve pulsations with periods larger than $\sim 1$~min, which might be expected for magnetic arches longer than 10~Mm. 
\new{The detection might be non-trivial, mostly because the regions contain several bright loops that intersect along the line of sight.} 

Here we describe the detection of pulsations from SDO/AIA images in regions with IRIS hot spots and we show their consistency with predictions by hydrodynamic loop modeling. We demonstrate that the observed QPP could be caused by a slow wave evolving in a closed magnetic loop and triggered by a short-time heating pulse deposited at the loop footpoints and in the corona. For the first time we link the observed properties of QPP with the properties of the underlying heating process, suggesting a new way of exploiting the QPP observed in the thermal emission for the diagnostics of the coronal heating function.

\section{Data analysis and modeling}
\label{sec:analysis}

\subsection{Observation}
\label{sec:obs}

We consider loops in an active region observed in the EUV with 
the Atmospheric Imaging Assembly \citep[AIA;][]{Lemen2012a} on board the Solar Dynamics Observatory (SDO) on 12 November 2015 starting at 1:35~UT and studied also in \cite{Reale2019a}.  The center of the field of view is XY=[-117.2", -329.6"]. \new{The loops brighten for a too short time in the 131~\AA\ channel to show a meaningful pulsation trend (as confirmed by hydrodynamic simulations), so here} we focus on the 94~\AA\ channel which includes highly ionized Fe\,{\sc xviii} line, most sensitive to plasma at temperatures in the range 6-8~MK \citep{Testa2012c,Boerner2014a}.  The data processing and analysis is the same as in \cite{Reale2019a}. The images have been preprocessed with the standard AIA software procedure and co-aligned, and we subtract the image just before the brightening of the hot structures (1:37.12~UT). \new{The dataset consists of 205 images with an average cadence of 12~s in a total time lapse of 30~min, between 01:35 and 02:05~UT, in which the subtracted image is not too far in time.}

\begin{figure}              
 \centering
   \subfigure[]
   {\includegraphics[width=9cm]{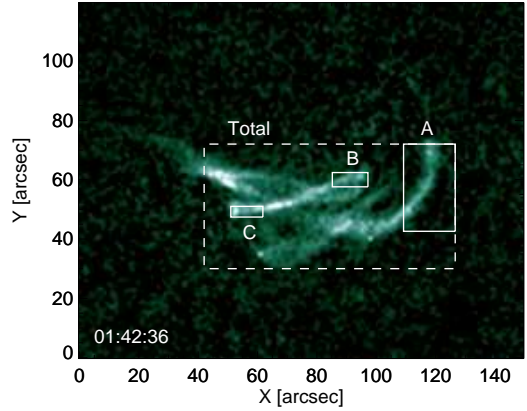}}
      \subfigure[]
   {\includegraphics[width=9cm]{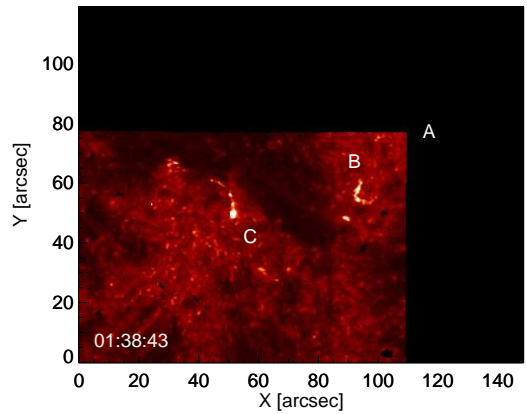}}
      \subfigure[]
   {\includegraphics[width=9cm]{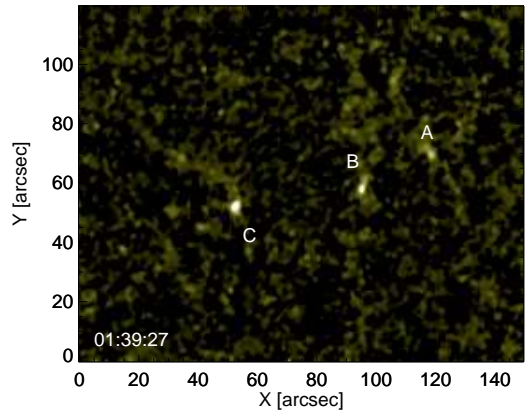}}
\caption{\footnotesize (a) Loop system  observed on 12 november 2015 in the AIA 94~\AA\ channel. An image just before the brightening (1:37:12~UT) was subtracted. \new{Light curves in Fig.~\ref{fig:obs_lc} were extracted from the boxes (Total,A,B,C). (b) Same region observed with IRIS in the Si\,{\sc iv} 1402~\AA\ line. The position of boxes A,B,C is marked. (c) Same as (b) but in the AIA 1600~\AA\ channel.}  } 
\label{fig:boxes}
\end{figure}

\begin{figure}              
 \centering
   {\includegraphics[width=14cm]{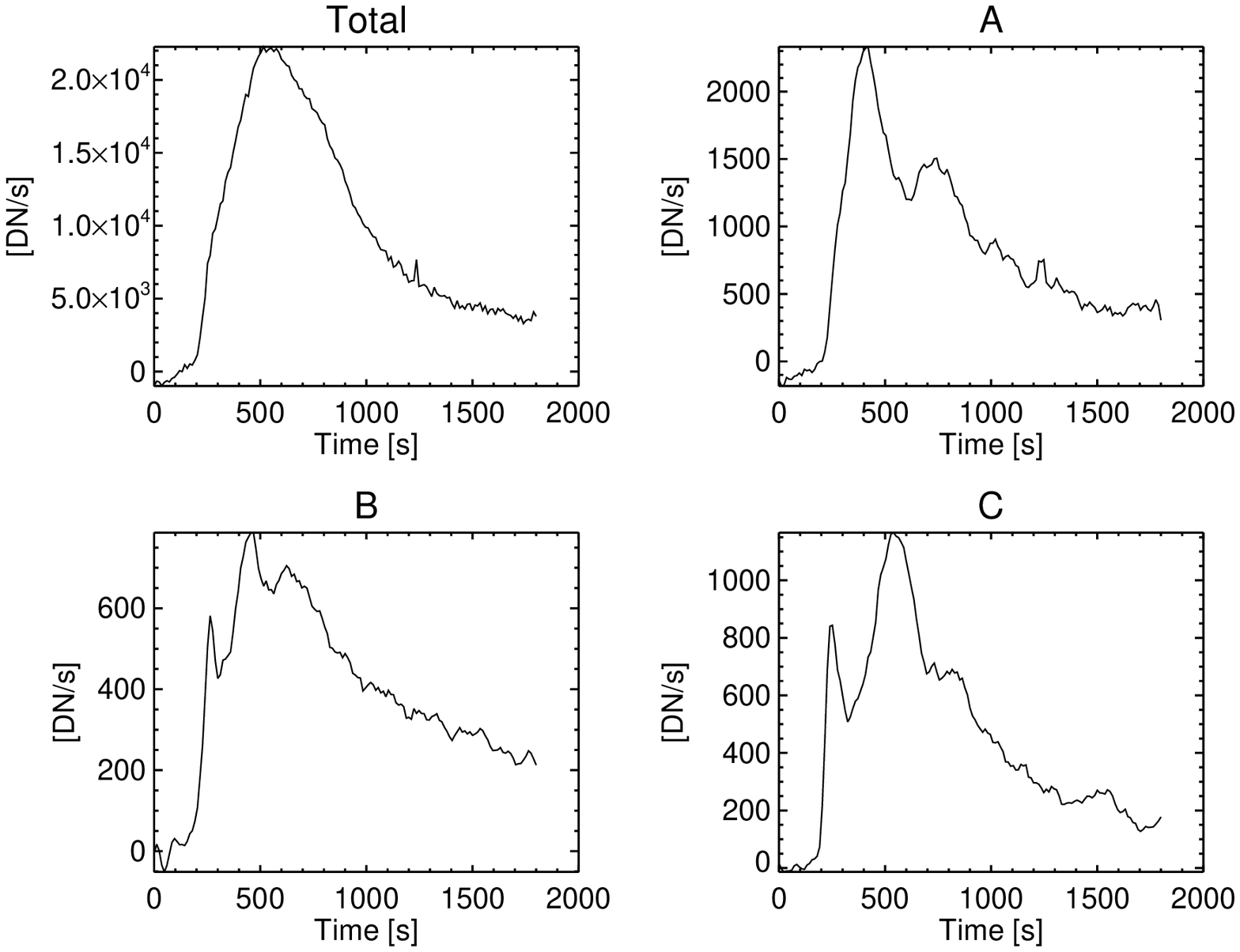}}
\caption{\footnotesize Light curves from the boxes in Fig.\ref{fig:boxes} (time since 01:35~UT).} 
\label{fig:obs_lc}
\end{figure}

We analyse the emission from the brightening region in Fig.\ref{fig:boxes}a, and in particular three boxes \new{A,B ,C}. The boxes are chosen so as to avoid bright overlapping structures. Box~A is taken along a long coronal arch, whose distance between the footpoints is $\sim 57$~Mm, for a maximum total length $\sim 90$~Mm if the loop were semicircular\citep{Reale2019a}. Boxes~B and C are at the extreme of a shorter transversal structure, whose distance between the footpoints is $\sim 32$~Mm, for a possible maximum length 50~Mm\citep{Reale2019a}. \new{In Figs.\ref{fig:boxes}b,\ref{fig:boxes}c the same region imaged by IRIS in the Si\,{\sc iv} 1402~\AA\ line and AIA 1600~\AA\ channel, respectively, shows bright spots at the footpoints of both loops.}

Figure~\ref{fig:obs_lc} shows the light curves related to this region in the 94~\AA\ channel after background subtraction, over \new{$\sim 30$~min}. The total light curve is integrated on the region \new{which contains the loops brightening in this channel and} shows a smooth flare-like evolution with a sharp rise, a well-defined peak approximately 5min after the brightening starts and a more gradual decay. This light curve envelopes the others. 

The light curves taken in the boxes are significantly different from the total one and share a common feature: all of them show well-defined large-amplitude pulsations. The light curve in box~A shows a first high peak at time $t \approx 550$~s, \neww{rapidly decaying to} a second smaller and smoother one ($t \approx 900$~s), and a third barely visible one. Both box B and C share another common feature: an initial similar and almost simultaneous ($t \approx 400$~s) spike. This spike is followed by a peak similar to the peak in box~A and by a third bump similar to the second bump in box~A. The amplitude of the main pulsations is 20-40\% of the total signal. The separation in time between the peaks, which can be taken as the period of the pulsations, is approximately 300~s for box~A, 200~s for box~B and 250-300~s for box~C. Pulsations are not detected everywhere in the region, mostly because different structures overlap along the line of sight.

\begin{figure}              
 \centering
  \subfigure[]
   {\includegraphics[width=8cm]{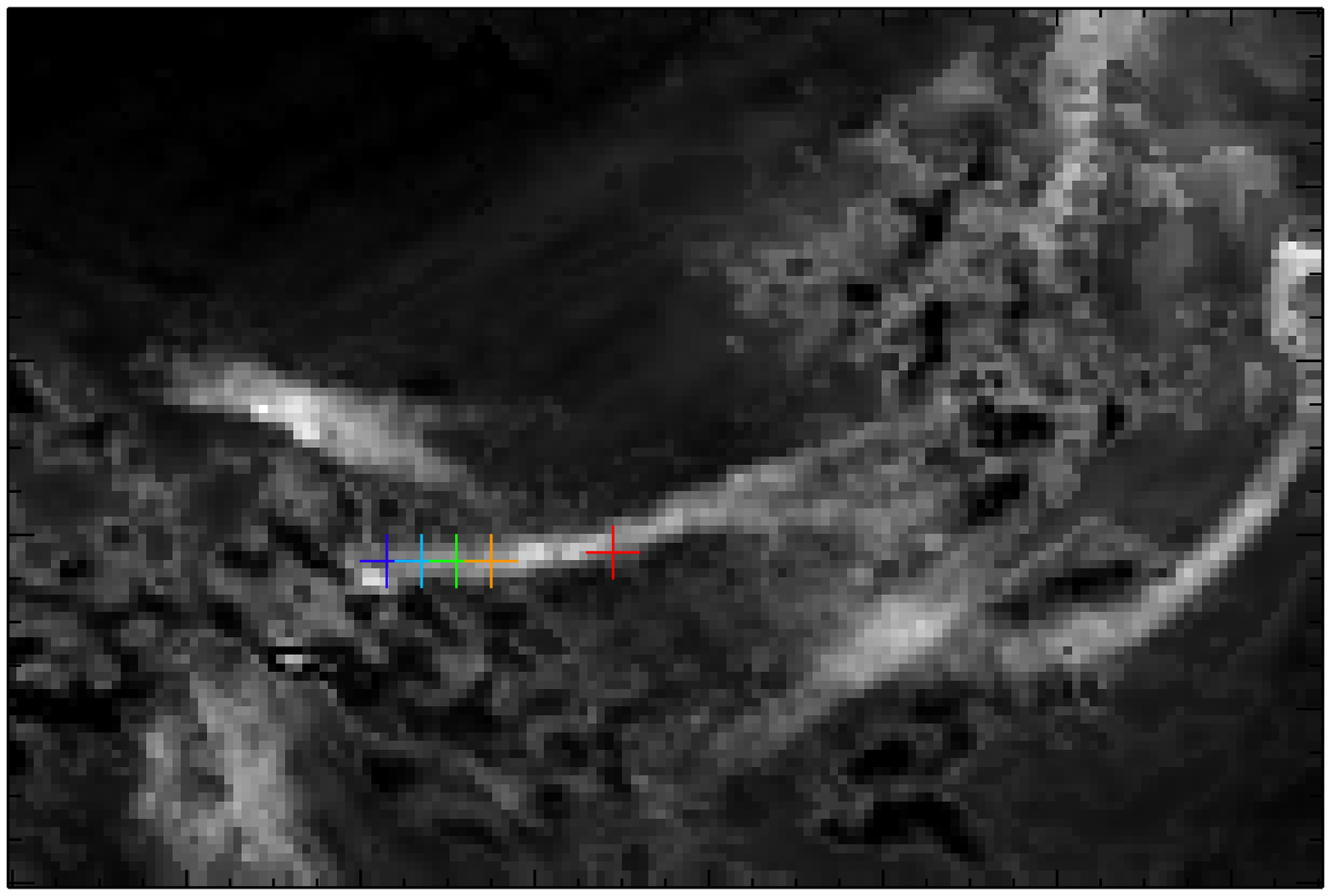}}
 \subfigure[]
   {\includegraphics[width=14cm]{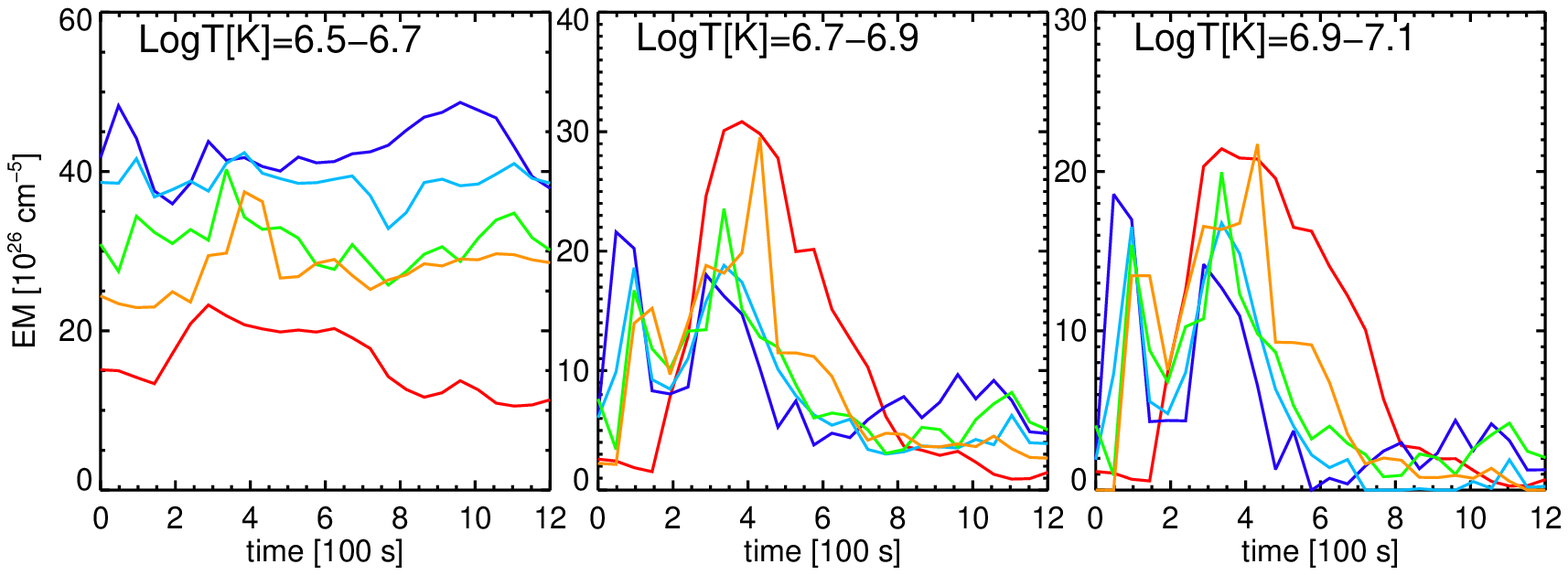}}
\caption{\footnotesize(a) Map of emission measure in the 6.7-6.9~$\log T[K]$ bin, at 01:41:54~UT. Four locations along one of the pulse-heated loops are marked (colored crosses). (b) Emission measure in \newb{3} temperature bins (increasing from left to right, as labelled in each panel) vs time for the four locations in panel~(a) with corresponding colors (blue to red, from the loop footpoints to larger heights along the loop). The reference time (t=0) is 01:37:54~UT. }
\label{fig:dem}
\end{figure}

The oscillations are also evident in the temporal evolution of the differential emission measure (DEM), which describes the temperature distribution of the coronal plasma along the line of sight. As detailed in a companion paper\citep{Reale2019a}, we applied the inversion method by \cite{Cheung2015a} to the timeseries of the 6 coronal AIA passband (94, 131, 171, 193, 211, 335~\AA) and obtained the DEM in each AIA pixel (no background subtraction is applied here).
Figure~\ref{fig:dem} shows the map of emission measure (EM) distribution in the 6.7-6.9 $\log T[K]$ bin, at 01:41:54~UT, and the temporal evolution of the emission measure in \newb{3} increasing-temperature bins, for 4 different locations along a heated loop (corresponding to box~C in Fig.\ref{fig:boxes}). At high temperatures ($\log T[K] \gtrsim 6.7$) the EM vs time shows pulsations. Excluding the curve for the point close to the apex (red) where other structures clearly overlap along the line of sight, we notice that the first pulsation shifts in time from the footpoint (blue) toward the apex (orange) by about 100~s (see Section~\ref{sec:discus}).

\subsection{Modeling}


\neww{We now show that pulsations similar to the observed ones are reproduced by hydrodynamic modeling of pulse-heated loops as in \cite{Reale2016a}.} We consider a standard loop model \neww{in the infinite magnetic field approximation}, which describes the time-dependent hydrodynamics of a compressible plasma confined in a closed coronal loop anchored to the photosphere \citep{Peres1982a,Betta1997a,Reale2016a}. \newb{The one-dimensional time-dependent hydrodynamic equations include the effect of gravity (for a curved flux tube), thermal conduction and compression viscosity \citep{Spitzer1962a}, radiative losses for an optically thin plasma, and an external heating input.} \neww{Under this approximation, the excited slow wave propagates strictly along the ambient magnetic field of an infinite strength, thus not perturbing the field and at the speed independent of it \citep{Goedbloed2004a,Priest2014a}.} \new{The model loop is assumed symmetric with respect to the apex. }

\begin{figure}              
 \centering
  \subfigure[]
   {\includegraphics[width=6cm]{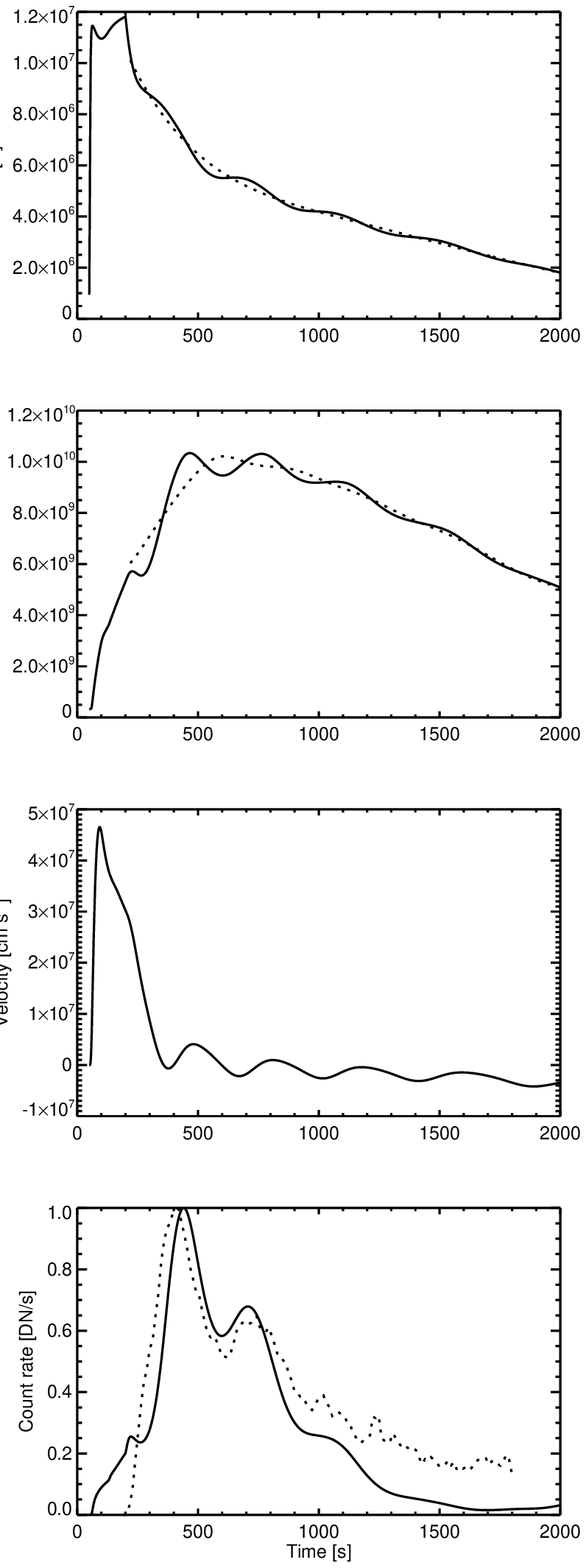}}
     \subfigure[]
   {\includegraphics[width=6cm]{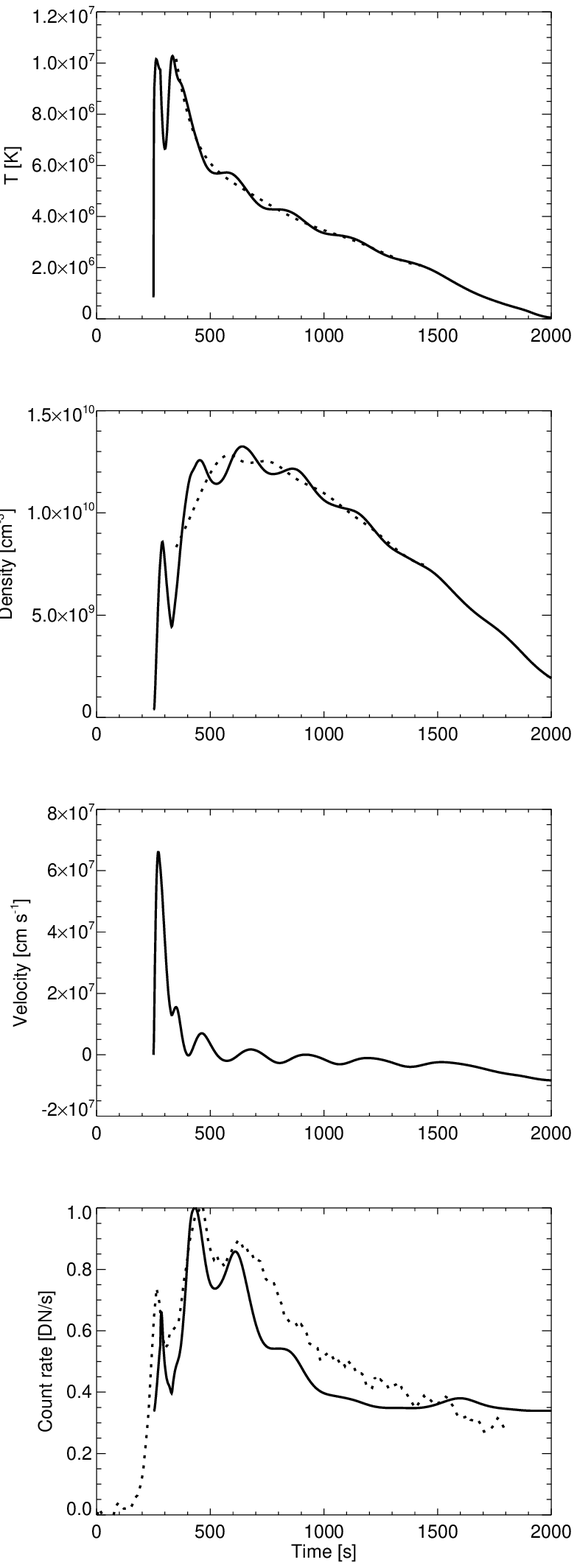}}
\caption{\footnotesize Results of hydrodynamic modeling of (a) a coronal loop heated with a pulse deposited uniformly in the loop and (b) a shorter loop heated with a pulse deposited at the footpoints (see text for details): from top to bottom, evolution of the average temperature, density, velocity, and normalized light curve (solid lines) synthesised in the AIA 94~\AA\ channel in low loop segment of 30~Mm (a) and 20~Mm (b), roughly corresponding to the loop leg enclosed in box~A (a) and box~B and~C (b) of Fig.\ref{fig:boxes}. \newb{The trends of the temperature and density variations used for estimate of $\gamma$ are also shown (dashed lines). The normalized light curves of box A and B (dashed lines) in Fig.~\ref{fig:obs_lc} are also shown for comparison. An offset of $\sim 50$\% of the maximum emission has been added to the model light curve in (b). }}
\label{fig:mod_top}
\end{figure}


Figure~\ref{fig:mod_top}a shows results of modeling of a loop symmetric with respect to its apex, with a half-length $L = 60$~Mm and heated by a single energy pulse with a duration of $\tau_H = 150$~s, an intensity of 0.2~erg~cm$^{-3}$~s$^{-1}$ and distributed uniformly along the loop. The total energy flux is $2.4 \times 10^9$~erg~cm$^{-2}$~s$^{-1}$. The loop is assumed initially at a maximum temperature of $\sim 1.5$~MK. Details about hydrodynamic loop modeling can be found in \cite{Reale2016a} and references therein. \new{The figure includes the evolution of the average temperature, density and velocity in the lower loop segment, 30~Mm long, roughly corresponding to the one enclosed in box A of Fig.~\ref{fig:boxes}}. The panels show the typical flare-like evolution with a very steep rise of the temperature $\sim 12$~MK, a more gradual rise of the density, due to plasma evaporation from the chromosphere \citep[e.g.,][]{Bradshaw2013a,Reale2014a}, to values $\sim 10^{10}$~cm$^{-3}$. After the heat pulse stops the temperature decreases very rapidly, and the density much more gradually, as expected \citep[e.g.,][]{Bradshaw2010a,Reale2014a}. From the maximum temperature the condition \citep{Reale2016a}:

\begin{equation}
\tau_H < \tau_s \sim 5 \frac{L_{Mm}}{\sqrt{0.1 ~  T_{MK}}} \approx 245~s
\label{eq:tau}
\end{equation}
\newb{(where $\tau_s$ is the return sound crossing time along the model half-loop, assuming a polytropic index $\gamma \sim 1$)} on the duration of the heat pulse confirms that the pulse is able to trigger pulsations inside the loop, which are clearly visible in the density of Fig.~\ref{fig:mod_top}a. \new{The 3 main pulsations are included in the time range 500-1500~s}, and the period is therefore $\sim 300-350$~s. The corresponding speed of the twin wavefronts is $\sim 350-400$~km/s. If we assume wavefronts at the sound speed \newd{of a fully ionized plasma $c_s = \sqrt{2 \gamma k_B T / m}$ where $k_B$ is the Boltzmann constant and $m$ is the average ion mass}, we can make a check of consistency about the temperature $T$ of the medium in which they propagate, under an assumption of the polytropic index $\gamma$.


\begin{figure}              
 \centering
   {\includegraphics[width=14cm]{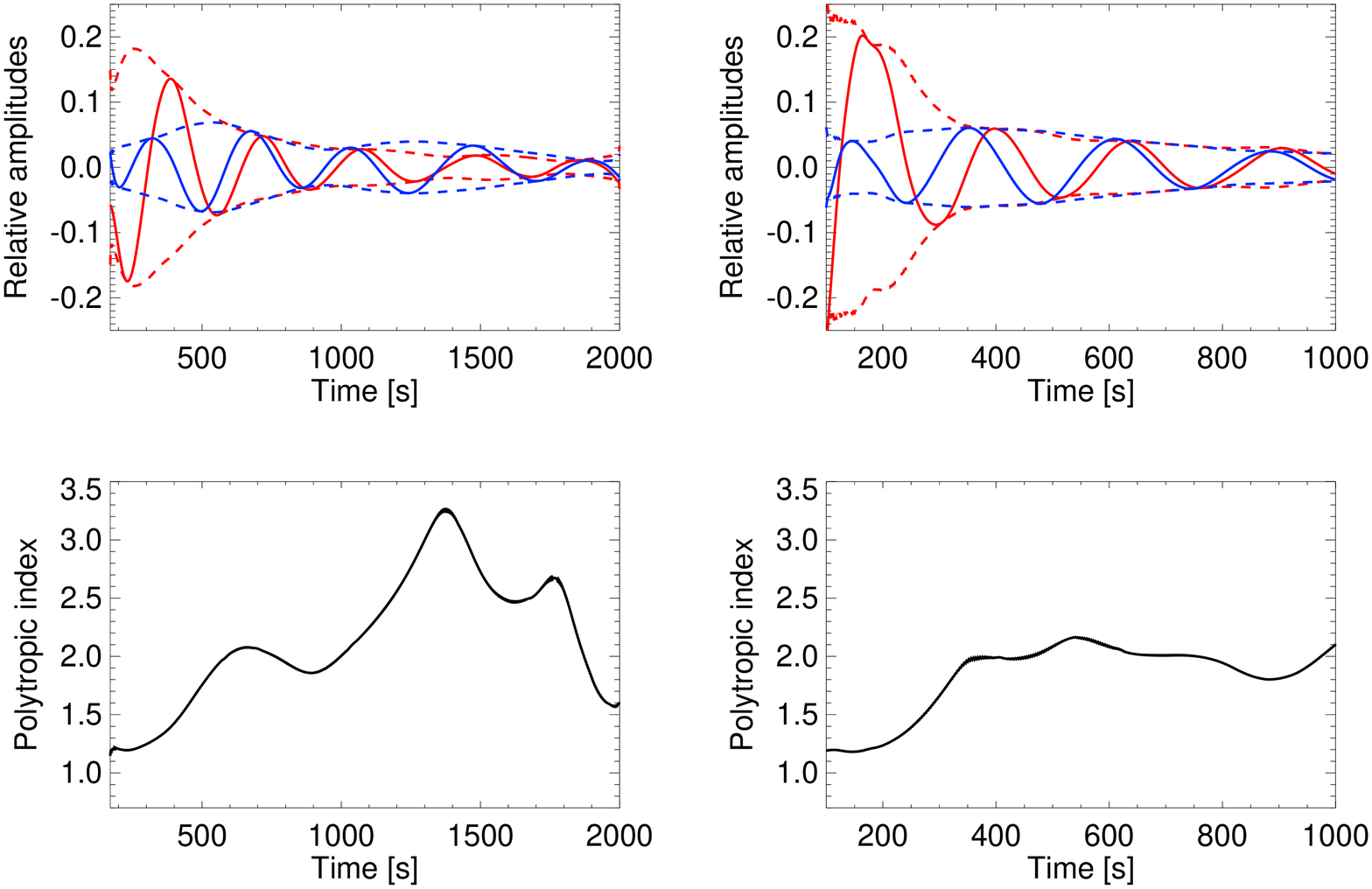}}
\caption{\footnotesize \newb{{\it Top:} Relative temperature (blue) and density (red) variations, normalised to the corresponding trends $T_0(t)$ and $n_0(t)$ (for the model as on left-hand and right-hand sides as in Fig.~\ref{fig:mod_top}), and their instantaneous amplitudes $A_T(t)$ and $A_n(t)$ obtained using the Hilbert transform. {\it Bottom:} Variation of the polytropic index with time, estimated from Eq.~(\ref{eq:gamma}).}} 
\label{fig:gamma}
\end{figure}

We can estimate the polytropic index $\gamma$ in our modelling results using the technique previously applied to a number of real observational data sets \citep[see e.g.,][]{Van-Doorsselaere2011a,Wang2015a}. It allows one to estimate $\gamma$ as
\begin{equation}\label{eq:gamma}
\gamma = \frac{A_T(t)}{A_n(t)} + 1,
\end{equation}
where $A_{T}(t)$ and $A_{n}(t)$ are the instantaneous relative amplitudes of the temperature and density variations, respectively, normalised to their corresponding trends $T_{0}(t)$ and $n_{0}(t)$ (see Fig.~\ref{fig:mod_top}). We determine $A_{T}(t)$ and $A_{n}(t)$ using the Hilbert transform of the corresponding temperature and density variations shown in Fig.~\ref{fig:gamma}. For both loop \newc{models}, the obtained values of the polytropic index $\gamma$ are seen to vary irregularly being about 1.2 (at the initial phase of the loop evolution with $T>8$\,MK), and from about 1.2 to approximately 5/3 (for the stage of a rapid cooling from 8 MK to 6 MK), and up to 2 and higher for later times ($T<5.5$\,MK). In order to interpret such a behaviour of $\gamma$, \newc{we point out that the derivation of Eq.(\ref{eq:gamma}) is based on the polytropic assumption, i.e., $p= K \rho^\gamma$ with small perturbations, independent of wave dissipation assumption, and that the estimate of the polytropic index using Eq.(\ref{eq:gamma}) includes all non-adiabatic thermohydrodynamic effects, that is why the measured values are beyond the expected range from thermal conduction alone (where $\gamma=1-5/3$) \citep[see Eqs.(5) in][]{Wang2015a}. For example, the wave-induced imbalance between heating and cooling processes, both present in our model, could also lead to the substantial modifications of $\gamma$ \citep{Zavershinskii2019a}.}
Nevertheless, at the initial stage of the loop evolution, the detected values of $\gamma \approx 1.2$ are consistent with the dominant thermal conduction in the loop evolution and support the assumption of approximately isothermal loop. 



\newb{If we consider $\gamma$ increasing from $\sim 1.2$ to $\sim 3$, we derive a temperature  decreasing \newe{from $T \sim 11$~MK to $T \sim 4$~MK}, broadly consistent with the temperature trend shown in Fig.~\ref{fig:mod_top}a}.
Figure~\ref{fig:mod_top}a shows also a very high speed, up to $\sim 500$~km/s, of the initial evaporation front, which decreases below 100~km/s already within $\sim 300$~s. 

\new{The bottom panel of Fig.~\ref{fig:mod_top}a shows the light curve synthesized from the model in the AIA 94~\AA\ channel} with the usual assumption on the radiative losses from an optically thin plasma and from the standard temperature-dependent channel sensitivity function, \newb{and the observed one for comparison}. \cite{Reale2016a} reports on the emission in the same channel but for a shorter loop ($L=25$~Mm).
The synthetic light curve clearly shows the expected pulsations.  
The light curve shows a good agreement with that measured in box~A of Fig.~\ref{fig:boxes}: both of them show a first higher peak and a second lower one, approximately with the same ratio of intensity, i.e., a similar pulsation amplitude. The period is also similar. \new{According to the model, the first larger bump is due to a combination of the density pulsation and of the rapid cooling across the channel temperature sensitivity range (Fig.~\ref{fig:mod_top}a). The small bump at time $t \sim 400$s is only barely visible in the light curve of box~A, possibly because of data noise.}

This light curve shows qualitative differences from those in boxes~B and~C, and, in particular, it is unable to reproduce the initial spike. Another hydrodynamic simulation provides a better agreement with the light curves in boxes~B and~C. 

\begin{figure}              
 \centering
   {\includegraphics[width=16cm]{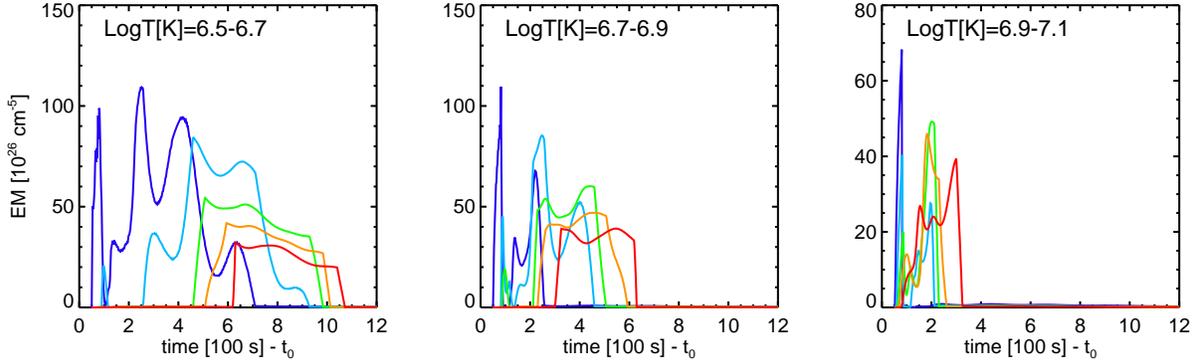}}
\caption{\footnotesize \newb{Emission measures vs time for 4 segments along the model loop in Fig.\ref{fig:mod_top}b respectively in the same temperature bins as in Fig.~\ref{fig:dem} for comparison. Time is shifted by $t_0 = 200$~s from that in Fig.~\ref{fig:mod_top}b. } }
\label{fig:dem_mod}
\end{figure}

Figure~\ref{fig:mod_top}b shows results from a model loop with half-length $L = 40$~Mm. \new{Twin heat pulses with a duration of $\tau_H = 30$~s are deposited symmetrically at the loop footpoints, as Gaussian functions} with $\sigma = 1$~Mm at a height of 3~Mm from each footpoint and with an intensity of 20~erg~cm$^{-3}$~s$^{-1}$, corresponding to a total flux of $\approx 10^{10}$~erg~cm$^{-2}$~s$^{-1}$. In this case, the pulse duration threshold is $\tau_H < 160$~s, so pulsations are triggered, and we see them again clearly in the density. Here the quantities are averaged over the lower 20~Mm of the loop, \new{roughly corresponding to boxes~B and C in Fig.~\ref{fig:boxes}}. Although overall similar, the evolution shows some differences from the previous model. The larger energy input drives a more effective plasma evaporation to a higher density ($\sim 1.3 \times 10^{10}$~cm$^{-3}$) and speed ($\sim 700$~km/s). The shorter loop corresponds to a faster evolution and to a shorter period of the pulsations ($\sim 200$~s), with a corresponding wavefront speed of $\sim 400$~km/s (Fig.~\ref{fig:mod_top}b). \newc{For this model we derive similar values of the sound speed and the $\gamma$ index as for the previous model (see Fig.~\ref{fig:gamma})}. Another distinctive feature is the \new{initial density spike, followed by smoother pulsations}. This is also due to the more effective initial evaporation front driven by this more concentrated heat pulse located much closer to the chromosphere.

The bottom panel of Fig.~\ref{fig:mod_top}b) shows the light curve in the 94~\AA\ channel  \new{from the loop segment of 20~Mm} of this other simulation, \newb{and the one observed in box~B for comparison. The light curve has a very good match with that in box~B one if we add an offset of $\sim 50$\% of the maximum model emission. The initial spike is similar} to those in the light curves of boxes~B and~C (Fig.~\ref{fig:obs_lc}). The next two pulsations are also very similar to those observed after the spike in the boxes. The oscillation period is around~200~s. We conclude that the presence of the spike is a signature of heat pulses concentrated at the footpoints. We have ascertained that \new{some important} parameters of this simulation are quite well constrained: a heat pulse significantly longer than 30~s broadens the initial spike; a less intense pulse (half) produces a second pulsation much higher than the third one. \newb{The offset might indicate the presence of contributions from other transient components not included in our simple single loop model.}

\newb{Finally, Fig.~\ref{fig:dem_mod} shows the evolution of the emission measure in the same temperature bins as in Fig.~\ref{fig:dem} for 4 segments along the model loop with footpoint heating (Fig.~\ref{fig:mod_top}b). A comparison with Fig.~\ref{fig:mod_top}b shows a good qualitative agreement for the central temperature bin 6.7-6.9 logT[K], and in particular regarding the timing of the peaks, and of the relative height of the peaks, except for the red curve as expected (Section~\ref{sec:obs}). The coolest temperature bin is not directly comparable because in the observed one the background contribution is substantial. In the hottest bin, although the timing of the segments is the same, the time scale of the model evolution appear significantly shorter than the observed one. The comparison in this bin confirms that other hot components are probably present and our the model cannot account for a multi-thermal structure across the loop.}

\section{Discussion} 
\label{sec:discus}

In this work we show the detection of quasi-periodic large-amplitude pulsations during transient and hot brightenings of a coronal loop system. The pulsations are detected in the legs of two distinct loops that intersect each other along the line of sight, and their amplitude is a large fraction ($\gtrsim 20$\%) of the total signal. This loop system is one of those where IRIS observed a hot spot in the UV Si\,{\sc iv} 1402~\AA\ line (Testa et al. 2019, in preparation) and where the related Doppler shift could be explained only with a heat pulse driven by a non-thermal electron beam \citep{Testa2014a,Polito2018a}. The pulsations have been modelled \neww{caused by slow magnetoacoustic wave fronts moving along} closed magnetic channels \citep{Reale2016a} which can only occur with a short ($\sim 1$~min) heat pulse (Eq.~\ref{eq:tau}). Thus the observed QPP confirm the impulsive nature of the heating in this loop system, and even provides constraints on its duration.

We performed hydrodynamic modeling of coronal loops with impulsive heating that is able to reproduce the most important features of the \neww{quasi-periodic pattern}, i.e., the amplitude and the period, together with the shape of the light curve and \neww{of the oscillation}. The model does not fit the observation in all the details, but this is rarely the case for any detailed time-dependent modeling. We cannot exclude that other explanations are possible, but our results are obtained with a simple assumption on heating in a \neww{single} closed loop, and are consistent with other independent evidence. As described in \cite{Reale2016a} the short heat pulse produces a pressure imbalance that drives a strong non-linear pressure front traveling along the loop and reflecting against the twin front from the other leg at the loop top. A low-order standing mode settles inside the magnetic tube which acts as a wave guide. The wave propagation is purely hydrodynamic, so we cannot account for finite plasma-$\beta$ and different MHD regimes \citep{Nakariakov2017a}. The compressive wave fronts propagate approximately at the sound speed, and can therefore be interpreted in the more general framework of the slow magnetosonic waves \citep{Nakariakov2005a,Kumar2013a,Kumar2015a,Wang2015a}.

\newb{Our model is the same as used in many other loop studies, including efficient thermal conduction \citep{Spitzer1962a}, in broad agreement} with other wave spectroscopic \citep{Van-Doorsselaere2011a} and imaging  \citep{Krishna-Prasad2018a} observations which found a polytropic index $\gamma \approx 1$, although a larger value was found in a flare observation \citep{Wang2015a}.

\newb{The observed and modelled light curves (Figs.~\ref{fig:mod_top}) show a good agreement, including the fact that the AIA 94~\AA\ channel detects only the hottest part of the loops evolution, i.e., the event peak and initial cooling, and the related initial oscillation periods, after which the plasma exits the range of temperature sensitivity of the channel. The detection of more pulsation periods would be more appropriate  
to address other issues on longer time scales, such as wave damping \citep{Ofman2002a,De-Moortel2003a,Kumar2016a}.}

We remark also that our loop model makes only a simple assumption on the impulsive heating function. Moreover, it includes a realistic transition region and chromosphere, and therefore the waves are not reflected at the lower boundary because of reflection boundary conditions but because they hit against the transition region.

Our model assumes that the loops are symmetric with respect to their apex \citep{Peres1982a}. This assumption looks reasonable because both footpoints of both loops are observed to brighten (Fig.~\ref{fig:boxes}). The slight asymmetries between the light curves of boxes~B and C in Fig.~\ref{fig:obs_lc} are probably due to slight differences in the heating deposition, which we do not address here. We also point out that the loop lengths estimated from the observation do not match the loop lengths assumed in the modeling. Although the measured values represented the starting point for modeling, the best model lengths are fine-tuned directly from forward modeling until a good match of the light curves is found, both regarding the time scales, and in particular the decay time \newd{of the light curves}, and the pulsation periods. The possible presence of magnetic twisting might explain why the measured lengths are smaller than the model lengths. According to \cite{Priest2014a}, the ratio of the model to the observed length gives the twisting in units of 2~$\pi$. The resulting values of 1.3 and 1.6 do not look unreasonable.

The model provides us with useful physical insight and with important constraints. First, it is an independent check which confirms the interpretation given for IRIS observations of UV hot spots \citep[][Testa et al. 2019, in preparation]{Testa2014a,Polito2018a}, and observed moss rapid variability observed with the high spatial resolution coronal imager Hi-C \citep{Testa2013a}. 
\neww{As mentioned above, the combination of density pulsations and rapid cooling shown in the first two rows of Figs.~\ref{fig:mod_top} determines the emission evolution in the bottom panels, and we see only a couple of well-defined pulsations, because of the narrow temperature sensitivity of the AIA 94~\AA\ channel \citep[see][for more details]{Reale2016a}. }

The fine details of the observation allow us to push the model to constrain the location of the heat pulses inside the closed loops. We find that the pulsation sequence of the loop structure whose footpoint at the footpoint of which IRIS Doppler shifted spectra are observed includes a spike that is well explained if the heat pulse is located at the loop footpoints. This is probably consistent with the presence of non-thermal electron beams \neww{generated from  reconnection of interacting magnetic structures and} hitting the chromosphere at the footpoints. The duration of the heating is typically not easily constrained for impulsive events \citep[e.g.,][]{Peres1987a,Reale2000a,Klimchuk2006a}, but in this case the model constrains the duration of the heat pulse to be around 30~s, in agreement with estimates derived from the modeling of the IRIS brightenings (Testa et al. 2019, in preparation). Also the intensity of the heating is usually largely unknown but in this case we find that $10^{10}$~erg~cm$^{-2}$~s$^{-1}$ is probably a very good reference value.
The loop footpoints brightenings are short-lived ($\leq 1$~min) in the AIA 1600~\AA\ channel (Fig.~\ref{fig:boxes}c), which is consistent with the short duration of the heat pulse\citep[e.g.,][]{Qiu2012a,Testa2014a}.

The emission measure reconstruction along the loop confirms the presence of the pulsations at high temperature ($\log T[K] \gtrsim 6.7$). \newb{In the limits of a single loop model which cannot account for thermal structuring across the loop, and for the brightening of other nearby structures, the model is able to reproduce the correct timing of the pulsations in the hot temperature bins and the details of them in the 6.7-6.9 $\log T[K]$ one, where the bulk of the brightening occurs (Fig.~\ref{fig:dem_mod}).} 
The high background probably prevents the detection of pulsations expected at later times at lower temperatures. In Fig.~\ref{fig:dem}b the first pulsation shifts by about 100~s from the leftmost to the third rightward point. This effect is probably real and marks the motion of the sloshing front. We measure an apparent speed of $\sim 75$~km/s. We see the loop projected on the disk and if we correct for the estimated real loop curvature we obtain a true speed of $\sim 200$~km/s, which becomes $\sim 320$~km/s if we include twisting factor (1.6). If we assume that the front moves at the average (isothermal) sound speed $c_s \sim 0.11~T^{1/2}$~km/s we obtain a temperature of $\sim 8$~MK, quite consistent with the loop model.


The pulsations in another (and longer) loop structure show some qualitative difference (no spike) and allow us to constrain that the heat pulse is probably deposited higher in the corona. This loop structure is transverse to, and probably interacting with, the other. The efficient thermal conduction in the corona makes difficult to constrain the location more in this case \citep[e.g.,][]{Reale2000a}, any location will produce similar results there. We have a constraint of the pulse duration which is longer in this case, about 2.5~min. 

We find then that in the same event we have heating deposited at loop footpoints and in the corona, with duration of 0.5 and 2.5~min, respectively.  This difference of heating location and duration in two nearby (probably interacting) structures within the same event broadens the scenario of impulsive heating deposition, and may have important implications within the framework of large-scale magnetic rearrangements and for the whole coronal heating theory.

In the end, the pulsations detected here are very strong evidence of the presence of impulsive heating inside a coronal loop system, \new{similar to studies of Doppler loop oscillations observed by SOHO/SUMER \citep{Wang2005a}. The predicted very high (to several hundreds km/s), hot but short-lived blue-shifts coming up from the chromosphere at the very beginning of the loop brightening shifts might be detected by future EUV spectrometers and support further} impulsive energy releases in active regions.

\acknowledgments{F.R., A.P., acknowledge support from  Italian Ministero dell'Istruzione, dell'Universit\`a e della Ricerca. PT acknowledges support by NASA grants NNX15AF50G and NNX15AF47G, and by contracts 8100002705 and SP02H1701R from Lockheed-Martin to SAO. D.Y.K. acknowledges support by the STFC consolidated grant ST/P000320/1. The authors thank the International Space Science Institute (ISSI) for their support and hospitality during the meetings of the ISSI team “New Diagnostics of Particle Acceleration in Solar Coronal Nanoflares from Chromospheric Observations and Modeling.” }
 



\begin{thebibliography}{}
\expandafter\ifx\csname natexlab\endcsname\relax\def\natexlab#1{#1}\fi

\bibitem[{Betta {et~al.}(1997)Betta, Peres, Reale, \& Serio}]{Betta1997a}
Betta, R., Peres, G., Reale, F., \& Serio, S. 1997, Astron. Astrophys. Suppl.
  Ser., 122, 585

\bibitem[{{Boerner} {et~al.}(2014){Boerner}, {Testa}, {Warren}, {Weber}, \&
  {Schrijver}}]{Boerner2014a}
{Boerner}, P.~F., {Testa}, P., {Warren}, H., {Weber}, M.~A., \& {Schrijver},
  C.~J. 2014, \solphys, 289, 2377

\bibitem[{Bradshaw \& Cargill(2010)}]{Bradshaw2010a}
Bradshaw, S.~J., \& Cargill, P.~J. 2010, Astrophys. J., 717, 163

\bibitem[{Bradshaw \& Cargill(2013)}]{Bradshaw2013a}
---. 2013, Astrophys. J., 770, 12

\bibitem[{{Cargill} {et~al.}(2015){Cargill}, {Warren}, \&
  {Bradshaw}}]{Cargill2015a}
{Cargill}, P.~J., {Warren}, H.~P., \& {Bradshaw}, S.~J. 2015, Philosophical
  Transactions of the Royal Society of London Series A, 373, 20140260

\bibitem[{{Cheung} {et~al.}(2015){Cheung}, {Boerner}, {Schrijver}, {Testa},
  {Chen}, {Peter}, \& {Malanushenko}}]{Cheung2015a}
{Cheung}, M.~C.~M., {Boerner}, P., {Schrijver}, C.~J., {et~al.} 2015, \apj,
  807, 143

\bibitem[{{Cho} {et~al.}(2016){Cho}, {Cho}, {Nakariakov}, {Kim}, \&
  {Kumar}}]{Cho2016a}
{Cho}, I.-H., {Cho}, K.-S., {Nakariakov}, V.~M., {Kim}, S., \& {Kumar}, P.
  2016, \apj, 830, 110

\bibitem[{{De Moortel} \& {Hood}(2003)}]{De-Moortel2003a}
{De Moortel}, I., \& {Hood}, A.~W. 2003, \aap, 408, 755

\bibitem[{{De Moortel} \& {Hood}(2004)}]{De-Moortel2004a}
---. 2004, \aap, 415, 705

\bibitem[{{De Moortel} {et~al.}(2004){De Moortel}, {Hood}, {Gerrard}, \&
  {Brooks}}]{De-Moortel2004b}
{De Moortel}, I., {Hood}, A.~W., {Gerrard}, C.~L., \& {Brooks}, S.~J. 2004,
  \aap, 425, 741

\bibitem[{De~Moortel \& Nakariakov(2012)}]{De-Moortel2012a}
De~Moortel, I., \& Nakariakov, V.~M. 2012, Royal Society of London
  Philosophical Transactions Series A, 370, 3193

\bibitem[{{Doyle} {et~al.}(2018){Doyle}, {Shetye}, {Antonova}, {Kolotkov},
  {Srivastava}, {Stangalini}, {Gupta}, {Avramova}, \&
  {Mathioudakis}}]{Doyle2018a}
{Doyle}, J.~G., {Shetye}, J., {Antonova}, A.~E., {et~al.} 2018, \mnras, 475,
  2842

\bibitem[{{Fang} {et~al.}(2015){Fang}, {Yuan}, {Van Doorsselaere}, {Keppens},
  \& {Xia}}]{Fang2015a}
{Fang}, X., {Yuan}, D., {Van Doorsselaere}, T., {Keppens}, R., \& {Xia}, C.
  2015, Astrophys. J., 813, 33

\bibitem[{{Goedbloed} \& {Poedts}(2004)}]{Goedbloed2004a}
{Goedbloed}, J.~P.~H., \& {Poedts}, S. 2004, {Principles of
  Magnetohydrodynamics}

\bibitem[{Hood {et~al.}(2009)Hood, Browning, \& van~der Linden}]{Hood2009b}
Hood, A.~W., Browning, P.~K., \& van~der Linden, R. A.~M. 2009, Astron.
  Astrophys., 506, 913

\bibitem[{{Inglis} {et~al.}(2016){Inglis}, {Ireland}, {Dennis}, {Hayes}, \&
  {Gallagher}}]{Inglis2016a}
{Inglis}, A.~R., {Ireland}, J., {Dennis}, B.~R., {Hayes}, L., \& {Gallagher},
  P. 2016, \apj, 833, 284

\bibitem[{Ionson(1978)}]{Ionson1978a}
Ionson, J. 1978, Astrophys. J., 226, 650

\bibitem[{{Kim} {et~al.}(2012){Kim}, {Nakariakov}, \& {Shibasaki}}]{Kim2012a}
{Kim}, S., {Nakariakov}, V.~M., \& {Shibasaki}, K. 2012, \apjl, 756, L36

\bibitem[{{Kliem} {et~al.}(2000){Kliem}, {Karlick{\'y}}, \&
  {Benz}}]{Kliem2000a}
{Kliem}, B., {Karlick{\'y}}, M., \& {Benz}, A.~O. 2000, \aap, 360, 715

\bibitem[{Klimchuk(2006)}]{Klimchuk2006a}
Klimchuk, J. 2006, Solar Phys., 234, 41

\bibitem[{{Kolotkov} {et~al.}(2018){Kolotkov}, {Pugh}, {Broomhall}, \&
  {Nakariakov}}]{Kolotkov2018a}
{Kolotkov}, D.~Y., {Pugh}, C.~E., {Broomhall}, A.-M., \& {Nakariakov}, V.~M.
  2018, \apjl, 858, L3

\bibitem[{Kopp \& Poletto(1993)}]{Kopp1993a}
Kopp, R., \& Poletto, G. 1993, Astrophys. J., 418, 496

\bibitem[{{Krishna Prasad} {et~al.}(2018){Krishna Prasad}, {Raes}, {Van
  Doorsselaere}, {Magyar}, \& {Jess}}]{Krishna-Prasad2018a}
{Krishna Prasad}, S., {Raes}, J.~O., {Van Doorsselaere}, T., {Magyar}, N., \&
  {Jess}, D.~B. 2018, \apj, 868, 149

\bibitem[{{Kumar} {et~al.}(2013){Kumar}, {Innes}, \& {Inhester}}]{Kumar2013a}
{Kumar}, P., {Innes}, D.~E., \& {Inhester}, B. 2013, \apj, 779, L7

\bibitem[{{Kumar} {et~al.}(2015){Kumar}, {Nakariakov}, \& {Cho}}]{Kumar2015a}
{Kumar}, P., {Nakariakov}, V.~M., \& {Cho}, K.-S. 2015, Astrophys. J., 804, 4

\bibitem[{{Kumar} {et~al.}(2016){Kumar}, {Nakariakov}, \& {Moon}}]{Kumar2016a}
{Kumar}, S., {Nakariakov}, V.~M., \& {Moon}, Y.~J. 2016, \apj, 824, 8

\bibitem[{{Kupriyanova} {et~al.}(2010){Kupriyanova}, {Melnikov}, {Nakariakov},
  \& {Shibasaki}}]{Kupriyanova2010a}
{Kupriyanova}, E.~G., {Melnikov}, V.~F., {Nakariakov}, V.~M., \& {Shibasaki},
  K. 2010, \solphys, 267, 329

\bibitem[{Lemen {et~al.}(2012)Lemen, Title, Akin, Boerner, Chou, Drake, Duncan,
  Edwards, Friedlaender, Heyman, Hurlburt, Katz, Kushner, Levay, Lindgren,
  Mathur, McFeaters, Mitchell, Rehse, Schrijver, Springer, Stern, Tarbell,
  W\"ulser, Wolfson, Yanari, Bookbinder, Cheimets, Caldwell, Deluca, Gates,
  Golub, Park, Podgorski, Bush, Scherrer, Gummin, Smith, Auker, Jerram, Pool,
  Soufli, Windt, Beardsley, Clapp, Lang, \& Waltham}]{Lemen2012a}
Lemen, J.~R., Title, A.~M., Akin, D.~J., {et~al.} 2012, Solar Phys., 275, 17

\bibitem[{McIntosh {et~al.}(2011)McIntosh, De~Pontieu, Carlsson, Hansteen,
  Boerner, \& Goossens}]{McIntosh2011a}
McIntosh, S.~W., De~Pontieu, B., Carlsson, M., {et~al.} 2011, Nature, 475, 477

\bibitem[{{McLaughlin} {et~al.}(2011){McLaughlin}, {Hood}, \& {de
  Moortel}}]{McLaughlin2011b}
{McLaughlin}, J.~A., {Hood}, A.~W., \& {de Moortel}, I. 2011, \ssr, 158, 205

\bibitem[{{McLaughlin} {et~al.}(2018){McLaughlin}, {Nakariakov}, {Dominique},
  {Jel{\'\i}nek}, \& {Takasao}}]{McLaughlin2018a}
{McLaughlin}, J.~A., {Nakariakov}, V.~M., {Dominique}, M., {Jel{\'\i}nek}, P.,
  \& {Takasao}, S. 2018, \ssr, 214, 45

\bibitem[{{Morton} {et~al.}(2019){Morton}, {Weberg}, \&
  {McLaughlin}}]{Morton2019a}
{Morton}, R.~J., {Weberg}, M.~J., \& {McLaughlin}, J.~A. 2019, Nature
  Astronomy, 196

\bibitem[{{Murray} {et~al.}(2009){Murray}, {van Driel-Gesztelyi}, \&
  {Baker}}]{Murray2009a}
{Murray}, M.~J., {van Driel-Gesztelyi}, L., \& {Baker}, D. 2009, \aap, 494, 329

\bibitem[{{Nakariakov} {et~al.}(2017){Nakariakov}, {Afanasyev}, {Kumar}, \&
  {Moon}}]{Nakariakov2017a}
{Nakariakov}, V.~M., {Afanasyev}, A.~N., {Kumar}, S., \& {Moon}, Y.~J. 2017,
  \apj, 849, 62

\bibitem[{{Nakariakov} {et~al.}(2006){Nakariakov}, {Foullon}, {Verwichte}, \&
  {Young}}]{Nakariakov2006a}
{Nakariakov}, V.~M., {Foullon}, C., {Verwichte}, E., \& {Young}, N.~P. 2006,
  \aap, 452, 343

\bibitem[{{Nakariakov} {et~al.}(2019){Nakariakov}, {Kosak}, {Kolotkov},
  {Anfinogentov}, {Kumar}, \& {Moon}}]{Nakariakov2019a}
{Nakariakov}, V.~M., {Kosak}, M.~K., {Kolotkov}, D.~Y., {et~al.} 2019, \apj,
  874, L1

\bibitem[{{Nakariakov} \& {Melnikov}(2009)}]{Nakariakov2009a}
{Nakariakov}, V.~M., \& {Melnikov}, V.~F. 2009, Space Science Reviews, 149, 119

\bibitem[{Nakariakov \& Verwichte(2005)}]{Nakariakov2005a}
Nakariakov, V.~M., \& Verwichte, E. 2005, Living Reviews in Solar Physics, 2, 3

\bibitem[{Ofman {et~al.}(1998)Ofman, Klimchuk, \& Davila}]{Ofman1998a}
Ofman, L., Klimchuk, J., \& Davila, J. 1998, Astrophys. J., 493, 474

\bibitem[{Ofman \& Wang(2002)}]{Ofman2002a}
Ofman, L., \& Wang, T. 2002, Astrophys. J. Lett., 580, L85

\bibitem[{Parker(1988)}]{Parker1988a}
Parker, E. 1988, Astrophys. J., 330, 474

\bibitem[{Peres {et~al.}(1987)Peres, Reale, Serio, \& Pallavicini}]{Peres1987a}
Peres, G., Reale, F., Serio, S., \& Pallavicini, R. 1987, Astrophys. J., 312,
  895

\bibitem[{Peres {et~al.}(1982)Peres, Serio, Vaiana, \& Rosner}]{Peres1982a}
Peres, G., Serio, S., Vaiana, G., \& Rosner, R. 1982, Astrophys. J., 252, 791

\bibitem[{{Polito} {et~al.}(2018){Polito}, {Testa}, {Allred}, {De Pontieu},
  {Carlsson}, {Pereira}, {Go{\v{s}}i{\'c}}, \& {Reale}}]{Polito2018a}
{Polito}, V., {Testa}, P., {Allred}, J., {et~al.} 2018, \apj, 856, 178

\bibitem[{Priest(2011)}]{Priest2011a}
Priest, E.~R. 2011, Journal of Atmospheric and Solar-Terrestrial Physics, 73,
  271

\bibitem[{{Priest}(2014)}]{Priest2014a}
{Priest}, E.~R. 2014, {Magnetohydrodynamics of the Sun} (Cambridge University
  Press)

\bibitem[{{Pugh} {et~al.}(2016){Pugh}, {Armstrong}, {Nakariakov}, \&
  {Broomhall}}]{Pugh2016a}
{Pugh}, C.~E., {Armstrong}, D.~J., {Nakariakov}, V.~M., \& {Broomhall}, A.-M.
  2016, \mnras, 459, 3659

\bibitem[{{Qiu} {et~al.}(2012){Qiu}, {Liu}, \& {Longcope}}]{Qiu2012a}
{Qiu}, J., {Liu}, W.-J., \& {Longcope}, D.~W. 2012, \apj, 752, 124

\bibitem[{{Reale}(2014)}]{Reale2014a}
{Reale}, F. 2014, Living Reviews in Solar Physics, 11, 4

\bibitem[{{Reale}(2016)}]{Reale2016a}
---. 2016, Astrophys. J. Lett., 826, L20

\bibitem[{{Reale} {et~al.}(2018){Reale}, {Lopez-Santiago}, {Flaccomio},
  {Petralia}, \& {Sciortino}}]{Reale2018a}
{Reale}, F., {Lopez-Santiago}, J., {Flaccomio}, E., {Petralia}, A., \&
  {Sciortino}, S. 2018, \apj, 856, 51

\bibitem[{Reale {et~al.}(2000)Reale, Peres, Serio, Betta, DeLuca, \&
  Golub}]{Reale2000a}
Reale, F., Peres, G., Serio, S., {et~al.} 2000, Astrophys. J., 535, 423

\bibitem[{{Reale} {et~al.}(2019){Reale}, {Testa}, {Petralia}, \&
  {Graham}}]{Reale2019a}
{Reale}, F., {Testa}, P., {Petralia}, A., \& {Graham}, D. 2019, \apj, submitted

\bibitem[{{Ruderman}(2013)}]{Ruderman2013a}
{Ruderman}, M.~S. 2013, \aap, 553, A23

\bibitem[{{Selwa} {et~al.}(2005){Selwa}, {Murawski}, \& {Solanki}}]{Selwa2005a}
{Selwa}, M., {Murawski}, K., \& {Solanki}, S.~K. 2005, \aap, 436, 701

\bibitem[{{Sim{\~o}es} {et~al.}(2015){Sim{\~o}es}, {Hudson}, \&
  {Fletcher}}]{Simoes2015a}
{Sim{\~o}es}, P.~J.~A., {Hudson}, H.~S., \& {Fletcher}, L. 2015, \solphys, 290,
  3625

\bibitem[{Spitzer(1962)}]{Spitzer1962a}
Spitzer, L. 1962, Interscience Tracts on Physics and Astronomy, Vol.~3, Physics
  of Fully Ionized Gases, 2nd edn. (New York: Interscience)

\bibitem[{{Svestka}(1994)}]{Svestka1994a}
{Svestka}, Z. 1994, \solphys, 152, 505

\bibitem[{Testa \& Reale(2012)}]{Testa2012c}
Testa, P., \& Reale, F. 2012, Astrophys. J. Lett., 750, L10

\bibitem[{Testa {et~al.}(2013)Testa, De~Pontieu, Mart{\'{\i}}nez-Sykora,
  DeLuca, Hansteen, Cirtain, Winebarger, Golub, Kobayashi, Korreck, Kuzin,
  Walsh, DeForest, Title, \& Weber}]{Testa2013a}
Testa, P., De~Pontieu, B., Mart{\'{\i}}nez-Sykora, J., {et~al.} 2013,
  Astrophys. J. Lett., 770, L1

\bibitem[{{Testa} {et~al.}(2014){Testa}, {De Pontieu}, {Allred}, {Carlsson},
  {Reale}, {Daw}, {Hansteen}, {Martinez-Sykora}, {Liu}, {DeLuca}, {Golub},
  {McKillop}, {Reeves}, {Saar}, {Tian}, {Lemen}, {Title}, {Boerner},
  {Hurlburt}, {Tarbell}, {Wuelser}, {Kleint}, {Kankelborg}, \&
  {Jaeggli}}]{Testa2014a}
{Testa}, P., {De Pontieu}, B., {Allred}, J., {et~al.} 2014, Science, 346,
  1255724

\bibitem[{{Thurgood} {et~al.}(2019){Thurgood}, {Pontin}, \&
  {McLaughlin}}]{Thurgood2019a}
{Thurgood}, J.~O., {Pontin}, D.~I., \& {McLaughlin}, J.~A. 2019, \aap, 621,
  A106

\bibitem[{van Ballegooijen {et~al.}(2011)van Ballegooijen, Asgari-Targhi,
  Cranmer, \& DeLuca}]{van-Ballegooijen2011a}
van Ballegooijen, A.~A., Asgari-Targhi, M., Cranmer, S.~R., \& DeLuca, E.~E.
  2011, Astrophys. J., 736, 3

\bibitem[{{Van Doorsselaere} {et~al.}(2016){Van Doorsselaere}, {Kupriyanova},
  \& {Yuan}}]{Van-Doorsselaere2016a}
{Van Doorsselaere}, T., {Kupriyanova}, E.~G., \& {Yuan}, D. 2016, Solar Phys.,
  291, 3143

\bibitem[{{Van Doorsselaere} {et~al.}(2011){Van Doorsselaere}, {Wardle}, {Del
  Zanna}, {Jansari}, {Verwichte}, \& {Nakariakov}}]{Van-Doorsselaere2011a}
{Van Doorsselaere}, T., {Wardle}, N., {Del Zanna}, G., {et~al.} 2011, \apjl,
  727, L32

\bibitem[{Vekstein \& Katsukawa(2000)}]{Vekstein2000a}
Vekstein, G., \& Katsukawa, Y. 2000, Astrophys. J., 541, 1096

\bibitem[{{Verwichte} {et~al.}(2008){Verwichte}, {Haynes}, {Arber}, \&
  {Brady}}]{Verwichte2008a}
{Verwichte}, E., {Haynes}, M., {Arber}, T.~D., \& {Brady}, C.~S. 2008, \apj,
  685, 1286

\bibitem[{{Wang}(2011)}]{Wang2011a}
{Wang}, T. 2011, \ssr, 158, 397

\bibitem[{Wang {et~al.}(2013)Wang, Ofman, \& Davila}]{Wang2013a}
Wang, T., Ofman, L., \& Davila, J.~M. 2013, Astrophys. J. Lett., 775, L23

\bibitem[{{Wang} {et~al.}(2015){Wang}, {Ofman}, {Sun}, {Provornikova}, \&
  {Davila}}]{Wang2015a}
{Wang}, T., {Ofman}, L., {Sun}, X., {Provornikova}, E., \& {Davila}, J.~M.
  2015, \apjl, 811, L13

\bibitem[{{Wang} {et~al.}(2018){Wang}, {Ofman}, {Sun}, {Solanki}, \&
  {Davila}}]{Wang2018a}
{Wang}, T., {Ofman}, L., {Sun}, X., {Solanki}, S.~K., \& {Davila}, J.~M. 2018,
  \apj, 860, 107

\bibitem[{{Wang} {et~al.}(2005){Wang}, {Solanki}, {Innes}, \&
  {Curdt}}]{Wang2005a}
{Wang}, T.~J., {Solanki}, S.~K., {Innes}, D.~E., \& {Curdt}, W. 2005, \aap,
  435, 753

\bibitem[{{Zavershinskii} {et~al.}(2019){Zavershinskii}, {Kolotkov},
  {Nakariakov}, {Molevich}, \& {Ryashchikov}}]{Zavershinskii2019a}
{Zavershinskii}, D.~I., {Kolotkov}, D.~Y., {Nakariakov}, V.~M., {Molevich},
  N.~E., \& {Ryashchikov}, D.~S. 2019, arXiv e-prints, arXiv:1907.08168

\end{thebibliography}


\end{document}